\documentclass[twocolumn]{aa} % for a referee version
\usepackage{graphicx}
\def\hii{\hbox{H{\sc ii}}}
\def\uchii{\hbox{UCH{\sc ii}}}
\def\lesssim{\mathrel{\hbox{\rlap{\hbox{\lower3pt\hbox{$\sim$}}}{\raise2pt\hbox{$<$}}}}}
\def\gtrsim{\mathrel{\hbox{\rlap{\hbox{\lower3pt\hbox{$\sim$}}}{\raise2pt\hbox{$>$}}}}}
\def\micron{\hbox{$\mu$m}}

\begin{document}

   \title{High Resolution Mid-infrared Imaging of W3(OH)}

   \author{B. Stecklum\inst{1}
             \and
         B. Brandl\inst{2}
             \and
        Th. Henning\inst{3}
	 \and
	I. Pascucci\inst{3}                       
%        S. Rinehart\inst{2}
         \and
        T.L. Hayward\inst{4}
         \and
        J.C. Wilson\inst{2}
% \fnmsep\thanks{Just to show the usage
%         of the elements in the author field}
          }

   \offprints{B. Stecklum}

   \institute{Th\"uringer Landessternwarte Tautenburg, Sternwarte 5,
                D--07778 Tautenburg\\
              email: stecklum@tls--tautenburg.de
         \and
          Center for Radiophysics \& Space Research, Cornell University, Ithaca, NY 14853
         \and
         Max-Planck-Institut f\"ur Astrophysik, K\"onigstuhl 17, D--69117 Heidelberg
%         \and
%           Astrophysikalisches Institut und Universit\"ats--Sternwarte,
%           Friedrich--Schiller--Universit\"at Jena, Schillerg\"a{\ss}chen 2-3,
%           D--07745 Jena     
          \and
%          Department of Astronomy, Cornell University, 226 Space Sciences Bldg., Ithaca, NY 14853
        Gemini Observatory, 670 N. A'ohoku Place, Hilo, HI 96720
             }

   \date{Received xxx; xxx}

   \abstract{
      We present results of our diffraction-limited mid-infrared imaging of the massive
   star-forming
   region W3(OH) with SpectroCam--10 on the 5-m Hale telescope at wavelengths of 8.8,
   11.7, and 17.9\,\micron{}.  The thermal emission from heated dust grains associated
   with the ultracompact \hii{} region W3(OH) is resolved and has a spatial extent
   of $\sim$2\arcsec{} in the N band. We did not detect the hot core source W3(H$_2$O)
%    by Keto et al. (\cite{keto92}).
   which implies the presence of at least 12 mag of
   extinction at 11.7\,\micron{} towards this source. These results together with
   other data were used to constrain the properties of W3(OH)
   and W3(H$_2$O) and their envelopes by modelling the thermal dust emission.
%   The cometary  \hii{} region 6\arcsec{}
%   north-east of W3(OH) is associated with weak diffuse thermal emission.
      \keywords{Stars: formation, circumstellar matter  -- ISM: dust, extinction
      -- Infrared: stars
               }
   }

   \maketitle
%
%________________________________________________________________

\section{Introduction}
The ``hot cores'' revealed in recent years by molecular line investigations are
small ($\lesssim 0.1$\,pc), very dense ($n\gtrsim 10^7$cm$^{-3}$), and hot ($> 100$\,K)
entities of giant molecular clouds (e.g., Cesaroni et al. \cite{cesa98}).  They are
considered to be the likely birthplaces of massive stars (e.g., Garay \& Lizano \cite{garay99},
Kurtz et al. \cite{kurtz00}).
Hot cores are frequently associated with ultracompact \hii{} regions (\uchii s) which
are more evolved and accessible to near-infrared (NIR) studies of their stellar
population (e.g., Feldt et al. \cite{feldt98}, Henning et al. \cite{henning01}).
% and radio wavelengths 
%; Wood \& Churchwell \cite{wood89},
% Kurtz et al. \cite{kurtz94}). 
Although the temperatures and sizes of hot cores suggest that they might be
conspicuous objects in the infrared sky, extinction in the mid-infrared (MIR,
N (10\,\micron{}) and Q (20\,\micron{}) bands) caused by the large column
densities ($N$(H$_2)\gtrsim 10^{23}$\,cm$^{-2}$) must not be neglected. 

Conclusions concerning the heating and the stellar content of hot cores 
have to be based on the knowledge
of their luminosity. This quantity is difficult to estimate since the immediate
neighbourhood of \uchii s often leads to source confusion, especially in the far-infrared (FIR)
range where these objects emit most of their energy and the angular resultion of spaceborn
observations is as yet poor. Radio  interferometry at mm/submm wavelengths
allow to  separate the dust continuum emission of the hot core from the free-free radiation 
of the adjacent \uchii. Such measurements were used to
constrain models of hot cores (Osorio et al. \cite{osorio99}). High-resolution ground-based
MIR observations, on the other hand, provide information on the spectral energy distribution
(SED) shortward of the peak flux. A corresponding study of the
IRc2 source in the Orion BN/KL complex (Gezari et al. \cite{gezari98}) illustrates their
importance. More recently, \cite{2002ApJ...564L.101D} were able to detect
MIR emission from the hot core G29.96$-$0.02 with a morphology similar to that of the
warm ammonia.
% On the other
% hand, the knowledge of the stellar population formed in hot cores and the individual
% stellar luminosities
% is required to discriminate between different models of high-mass star formation.
% For this purpose, 
We performed high resolution MIR imaging of hot cores, including W3(H$_2$O) and
the neighbouring \uchii{} W3(OH), in order to measure their flux densities or to provide
at least upper limits. While results for G10.47+0.03 will be the subject of a forthcoming
paper (Pascucci et al., in prep.), we present here our findings for W3(H$_2$O) and the \uchii{} W3(OH).

% A recent example for the revision of luminosities by means of high-resolution
% thermal infrared imaging is that of IRc2 in the Orion BN/KL complex (Gezari et
% al. \cite{gezari98}).

The \uchii{} W3(OH) is very well-studied in the radio domain by continuum and molecular line
observations. It is located at the distance of
2.2\,kpc (Humphreys \cite{hum78}) and harbours numerous OH masers.
The hot core W3(H$_2$O), also known
as Turner-Welch object (TW, Turner \& Welch \cite{tw84}), is situated $\sim6$\,\arcsec{} east
of W3(OH). %, thought to be
% responsible for the internal
% heating of the hot core W3(H$_2$O). 
This enigmatic source shows an outflow traced by the
proper motion of H$_2$O masers (Alcolea et al. \cite{alco92})
and is associated with a double-sided radio continuum jet, presumably of synchrotron nature
(Reid et al. \cite{reid95}). Recent interferometric imaging at 220\,GHz by Wyrowski et al. (\cite{wyrowski99}) revealed
another % heating 
source in the immediate neighbourhood of W3(H$_2$O), suggesting that the
region harbours a cluster of protostars.
The thermal infrared emission from W3(OH) has been studied by Keto et al. (\cite{keto92})
using one of the first MIR array cameras (Arens et al. \cite{arens87}).
% at the Wyoming Infrared Observatory. 
Keto et al. claimed the detection of W3(H$_2$O) at the wavelength of 12.2\,\micron{}
with a flux density of $45 \pm 10$\,mJy. 
W3(H$_2$O) and W3(OH) were the subject of a continuum and molecular line study of
\cite{2000ApJ...537..283V}. Since their results are based on single-dish
data which do not resolve the two objects, they cannot be directly compared to
our model presented below.

%__________________________________________________________________

\section{Observations}
The observations were performed using SpectroCam--10, the Cornell-built 8--13~$\mu$m
spectrograph/camera (Hayward et al. \cite{hayward93}) on the 5-m Hale
telescope\footnote{Observations at the Palomar Observatory
were made as part of a continuing collaborative agreement between the California
Institute of Technology and Cornell University.}. SpectroCam's detector is a
Rockwell 128$\times$128 Si:As BIB array. In camera mode, the pixel scale is
0\farcs25 with a circular field of view of 16\arcsec{}.
We applied filters with central wavelengths and bandwidths (in parentheses) of
8.8 (1.0), 11.7 (1.0), and 17.9 (0.5)\,\micron. 
The imaging observations
were performed on 1998 December 27th using the common chopping/nodding technique
with a chopper throw of 20\arcsec{} in north-south direction. The wavelet filtering
algorithm of Pantin \& Starck (\cite{pantin96}) was applied to the images. This
algorithm is useful for recovering faint extended emission otherwise hidden in the noise.
The high dynamic range of the individual frames allowed the application of a
shift-and-add procedure which includes resampling,
yielding a final pixel scale of 0\farcs125. To enlarge our field of view, we observed
two positions at 11.7\,\micron{}, the first centered on W3(OH) and the second offset the
east by 10\arcsec{}. Lastly, the frames were mosaicked.

The astrometry is based on the radio position
of W3(OH) which is assumed to coincide with the peak of the infrared emission.

The flux calibration was performed using $\alpha$~Tau as reference which was measured
at about the same airmass (1.2) as the target. For this purpose, the photometric
zero points from Cohen et al. (\cite{cohen92}) and the photometry from Cohen et al.
(\cite{cohen95}) for $\alpha$~Tau were used. The observations were performed during
photometric conditions, with an internal photometric error of less than 2\% as estimated
from the flux variation of the standard star. The 3$\sigma$ sensitivities (mJy/beam)
for the detection of point sources 
% (mJy/$\sq$\arcsec) 
in the final
images are as follows: % 5 (8.8\,\micron{}), 10 (11.7\,\micron{}), and 160 (17.9\,\micron{}).
4 (8.8\,\micron{}), 6 (11.7\,\micron{}), and 93 (17.9\,\micron{}).

% zero points           alpha Tau mags
% F(8.8) =46.9 Jy       -2.95
% F(11.7)=28.6 Jy       -3.07
% F(17.9)=12.2 Jy       -3.08 (Q)
% 

\section{Results}

\begin{figure}
        \resizebox{\hsize}{!}{\includegraphics{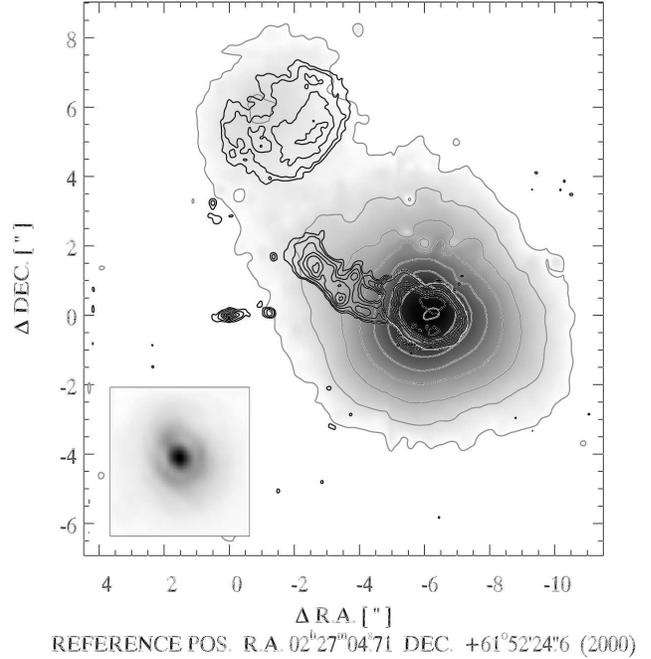}}
        \caption[]{11.7\,\micron{} image of W3(OH) with contours of the 8.4\,GHz
        radio continuum emission (black lines) from Wilner at al. (\cite{wilner99}).
        The grey contours delineate the 11.7\,\micron{} emission and are spaced by
        a factor of 2.5 starting at 3\,$\sigma$ (10\,mJy/$\sq$\arcsec).
        %The peak flux density is 7.7\,mJy/$\sq$\arcsec.
        W3(H$_2$O) is the elongated radio source at the reference position. The
        lower-left insert shows the 11.7\,\micron{} image of the reference star $\alpha$~Tau
        using a logarithmic brightness scale.}
        \label{Fig1}
   \end{figure}

The prime results of our observations are contained in Fig.\ref{Fig1} which shows
the 11.7\,\micron{} image together with the 8.4\,GHz radio continuum contours from Wilner at al.
(\cite{wilner99}, beam size 0\farcs2). 
% The high spatial resolution of the observations is obvious
% from the diffraction-limited 11.7\,\micron{} image of the reference star $\alpha$~Tau
% (Fig. \ref{Fig2}).
The two infrared sources in Fig.\ref{Fig1} correspond to W3(OH) and the cometary
\uchii{} situated north-east of it.
A comparison with $\alpha$~Tau
(Fig. \ref{Fig1} insert) shows that W3(OH) is spatially well resolved in our diffraction-limited
11.7\,\micron{} image (beam size 0\farcs6).
The lowest contour line of the infrared emission
represents the 3\,$\sigma$ detection limit. Clearly, the flux from W3(H$_2$O)
is below our sensitivity limit. This result is in contradiction to the infrared
detection of W3(H$_2$O) claimed by Keto et al. (\cite{keto92}).
We note that their conclusion is doubtful since the emission
they associate with W3(H$_2$O) peaks only 4\arcsec{} instead of 6\arcsec{} east of W3(OH)
(see their Fig.1). Furthermore, their total flux of 45$\pm$10\,mJy
contradicts with the fact that the emission well exceeds their lowest contour
level of 100\,mJy/$\sq$\arcsec. Thus, we conclude that they did not detect W3(H$_2$O)
as well, and were confused by the north-eastern cometary \uchii.

From Fig. \ref{Fig1} it is obvious that there is a good overall correspondence between the
MIR and the 8.4\,GHz radio
continuum emission. However, there are certain features which are different in both maps.
The north-eastern trail from W3(OH) is more confined in the radio continuum and only
marginally indicated by the infrared contours. The peak of the radio emission from the
north-eastern \uchii{} is closer towards W3(OH) than its infrared maximum. This is
presumably caused by the extinction of a dust lane stretching from W3(H$_2$O) to the
north-west which can be seen in  the C$^{17}$O(1-0) map of Wyrowski et al. (\cite{wyrowski97}).

%----------------------------------------------------
%______________________________________________________________

For the northeastern \uchii{} the following fluxes (within a 3\arcsec{}
aperture) were derived at the three wavelengths: 0.3, 0.6, and 6.1\,Jy,
respectively.

   \begin{table}
      \caption[]{SpectroCam--10, IRAS-LRS, and MSX flux densities of W3(OH)}
         \label{prop}
         \begin{tabular}{cccc}
            \hline
            \noalign{\smallskip}
            Wavelength  & Peak flux & Total flux   & FWHM  \\
            ~[\micron] &  [Jy]/$\sq$\arcsec & [Jy] &  [\arcsec] \\
            \noalign{\smallskip} 
            \hline
            \noalign{\smallskip}
                8.8   & 2.6  & 6.5   & 1.96 \\
                11.7  & 7.7  & 22.4  & 2.08 \\
                17.9  & 19.5 & 93.0   & 2.88 \\
            \noalign{\smallskip}
%            \hline
%         \end{tabular}
%      \caption[]{IRAS-LRS and MSX-SPIRIT~III flux densities}
%
%         \label{comp}
%         \begin{tabular}{cccc}
            \hline
            \noalign{\smallskip}
            IRAS-LRS  &  Total flux   & MSX  & Total flux  \\
            ~[\micron] &  [Jy] & [\micron] & [Jy] \\
            \noalign{\smallskip} 
            \hline
            \noalign{\smallskip}
                8.8   & 24.9  & 8.28   &  11.1\\
                11.7  & 24.1  & 12.13  &  36.7\\
                17.9  & 89.1  & 14.65  &  77.1\\
                      &       & 21.41  & 332.6\\
            \noalign{\smallskip}
            \hline
         \end{tabular}         
   \end{table}

\section{Discussion}
\subsection{The \uchii{} W3(OH)}
The FWHM of the MIR emission from W3(OH) was derived from Gaussian fits, taking the
size of the diffraction-limited beam into account. The sizes and flux densities
of W3(OH) are listed in Tab. \ref{prop}. The fluxes are based on an
aperture of 4\arcsec{} diameter. 
The extent of the MIR emission originating from the thermal emission of heated dust
grains exceeds that of the
8.4\,GHz radio continuum (FWHM 1\farcs52), indicating that
the warm dust is more extended than the ionized gas. The variation
of the angular size in dependence on wavelength can be approximated as
FWHM($\lambda$)\,$\sim \lambda^{0.6\pm0.2}$ and results from the decline
of the temperature with increasing distance from the heating star(s).

The comparison of our flux densities of W3(OH) with other estimates allows conclusions
on the influence of different beam sizes. For this purpose, we retrieved the IRAS-LRS
spectrum, identified W3(OH) in the MSX point source catalog (Egan et al. \cite{egan99}),
% MSX coords and fluxes in reduction log file sc10_98.reduct
and retrieved an ISO-LWS spectrum from the data archive.

%   \begin{table}
%      \caption[]{IRAS-LRS and MSX flux densities}
%         \label{comp}
%         \begin{tabular}{cccc}
%            \hline
%            \noalign{\smallskip}
%            IRAS-LRS  &  Total flux   & MSX isophotal wavelength & Total flux  \\
%            ~[\micron] &  [Jy] & [\micron] & [Jy] \\
%            \noalign{\smallskip} 
%            \hline
%            \noalign{\smallskip}
%                8.8   & 24.9  & 8.28   &  11.1\\
%                11.7  & 24.1  & 12.13  &  36.7\\
%                17.9  & 89.1  & 14.65  &  77.1\\
%                      &       & 21.41  & 332.6\\
%            \noalign{\smallskip}
%            \hline
%         \end{tabular}
%   \end{table}

The LRS spectrum was integrated according to the applied passbands. It is obvious
that the 8.8\,\micron{} fluxes given in Tab.\ref{prop} considerably exceed our value. This
can be explained by ubiquitous emission attributed to Polycyclic Aromatic Hydrocarbons
(PAHs) surrounding the \uchii{} which strongly contributes to the flux in the large
apertures of IRAS and MSX. Pronounced 7.7 and 8.6\,\micron{} PAH bands can be
misleading in ground-based derivations of the optical
depth of the 9.7\,\micron{} silicate feature (e.g., Roelfsema et al. \cite{roelf96}).
% The spectral index between 12.13 and 21.41\,\micron as derived from the MSX fluxes is
% about 3.9. Using this spectral index to extrapolate the 14.65\,\micron{} MSX flux to
% 17.9\,\micron{} yields 168\,Jy...

% W3(OH) and W3(H_2O) could be resolved with HAWC, perhaps also FORCAST on SOFIA
% ISO-LWS bolometric flux is 5.18e-20 W/cm-2, this yields a luminosity of 8x10^4 L_sun

%------------------------------------------------------
%   \begin{figure}[t]
%        \resizebox{\hsize}{!}{\includegraphics{W3_ISO.ps}}
%        \caption[]{ISO-LWS spectrum of W3(OH). The spectrum between 42 and 140\,\micron{}
%        can well be fitted by a modified black-body (35\,K and $\kappa(\nu)\sim\nu^2$,
%        dashed line).
%        The excess emission beyond 140\,\micron{} presumably originates from W3(H$_2$O).}
%        \label{Fig3}
%    \end{figure}
%______________________________________________________________

\subsection{The hot core W3(H$_2$O)}
% assuming Teff 150K and a size of the emitting region of 1 arcsec in diameter
% we expect the following fluxes from each source 2370, 2760, 2400 Jy.
% or surface brightnesses 1860, 2170, 1880 Jy per square arcsec
% comparing this to the 3 sigma detection limits yields a lower limit to the
% 11.7 micron extinction of 12.1mag
Our attempt to detect  W3(H$_2$O) in the IR was stimulated by the
presence of an outflow. Generally, IR emission can escape in outflow
lobes primarily due to scattering  (e.g., Fischer et al. \cite{fisch96}).
An example is NGC6334~I(N),
a presumed high-mass Class 0 object, for which
Sandell (Sandell \cite{sandell00}) rendered the detection of IR emission
impossible because of the extremely high extinction derived from
mm/submm maps. However, this source is associated with NIR emission 
(Tapia et al. \cite{tapia96}, Megeath \& Tieftrunk \cite{megeath99}) obviously
originating from the blue-shifted lobe of its outflow.

The flux densities from W3(H$_2$O) in the absence of any intervening absorbing matter
can be estimated from the temperature map
given by Wyrowski et al. (\cite{wyrowski97}). The expected peak surface
brightness amounts to 2170\,Jy/$\sq$\arcsec{} at 11.7\,{\micron}. Together with
our 3$\sigma$ sensitivity, this yields a lower limit to the extinction at this
wavelength of 12\,mag. Our failure to detect this source is consistent with the
high molecular hydrogen column densities of 1\dots3.5$\times$10$^{24}$\,cm$^{-2}$
inferred from molecular line and continuum investigations (Turner \& Welch \cite{tw84},
Wyrowski et al. \cite{wyrowski97}). It suggests that the molecular outflow of
W3(H$_2$O) is very young, i.e. did not fully penetrate the hot core yet, and,
in addition, might be in the plane of the sky. In fact,
Fig.\,1 from Wyrowski et al. (\cite{wyrowski99}) shows that the jet is confined
to the region of the hot core. The moderate expansion velocity of the H$_2$O masers
of 20\,km\,s$^{-1}$ (Alcolea et al. \cite{alco92})  implies
a dynamical timescale of only 500\,yr which is consistent with the upper limit
on proper motions of the radio jet of 150\,km\,s$^{-1}$ (Wilner et al. \cite{wilner99}).
These velocities are low compared to those of thermal radio jets (Anglada \cite{anglada96})
% N.B. der "synchrotron" jet kann sehr wohl auch ein thermischer sein ! 
and indicate that the outflow is presumably slowed-down by the high-density environment.

%other items:
%- consider the accretion case and also the possibility that any energy released by
%        stellar collisions will be thermalized at these high densities

\section{Modelling the dust emission}
One of the questions concerning W3(OH) and W3(H$_2$O) is related to their individual
luminosities which directly translates to the nature of the internal heating sources.
Although it is reasonable to assume that the internal heating is due to stars of
intermediate or high mass, no direct confirmation of their presence exists. The luminosity
might also, at  least partly, result from accretion. The high column densities toward
W3(H$_2$O) render it very difficult to figure out whether the mass accretion is due to
infall from a circumstellar disk or stellar mergers like in the scenario of Bonnell et
al. (\cite{bonnell98}). The high optical depths will lead to a thermalization of the released
energy irrespective of the accretion mechanism.

The derivation of the individual luminosities has to be based
on the decomposition of the SED. While our measurements provide
constraints for the MIR, the lack of spatial resolution
in the FIR does not permit to separate both components. 
An upper limit of $\sim$1800\,Jy on the 50\,\micron{} flux of W3(H$_2$O) has
been established by Campbell et al. (\cite{campbell89}) from KAO scans. 
% Future
% instruments like HAWC on SOFIA will improve this situation. 
In the mm/submm range, aperture
synthesis measurements
show that the 1.3\,mm emission from W3(H$_2$O) is due to dust radiation while
free-free emission dominates the flux of W3(OH) at this wavelength (Wyrowski et al.
\cite{wyrowski99}). The ISO-LWS spectrum %(Fig.\ref{Fig3}) shows that the SED 
between 42 and 140\,\micron{}
can well be fitted by a modified black-body (35\,K and $\kappa(\nu)\sim\nu^2$).
The association of W3(H$_2$O) with cold dust suggests that the MIR/FIR excess
in the ISO spectrum for $\lambda\,>\,140\,\micron$ is presumably due to the hot core. 
% The luminosity derived from the
% ISO-LWS spectrum is $8\times10^4\,L_{\sun}$.

\begin{figure}[t]
        \resizebox{\hsize}{!}{\includegraphics{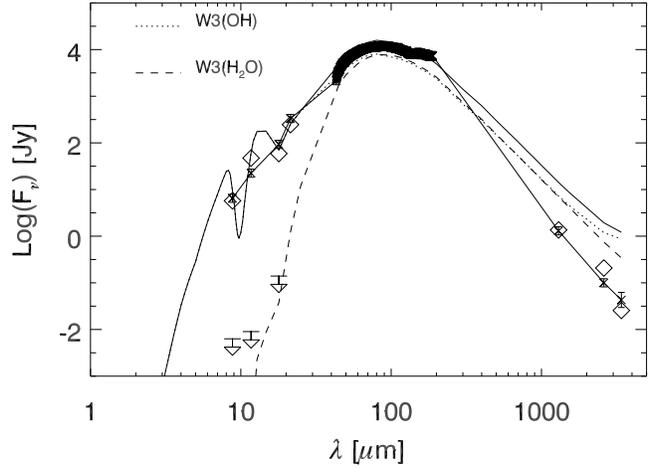}}
        \caption[]{SEDs of the individual sources 
	(W3(OH) -- dotted line,  W3(H$_2$O) -- dashed line)
	and of their sum (solid line). Asterisks with error bars indicate
        the measured flux densities. Diamonds represent the model fluxes
        for the observational beam sizes. Arrows mark the detection limits
        for W3(H$_2$O).}
        \label{fig:sed}
    \end{figure}

We modelled the thermal dust radiation from W3(OH) and W3(H$_2$O) in order to figure
out how much the hot core contributes to the FIR excess seen in the ISO-LWS spectrum and
which sensitivities are required for its detection in the MIR.
The input SED is based on fluxes coming from this work, from 
MSX point source catalogue at 21.4\,\micron{} (see Tab.~\ref{prop}), from the IRAS-LSR
and ISO-LWS spectra, and from
Wyrowski et al. \cite{wyrowski97} as well as Wilner et al. \cite{wilner95} for the mm
wavelenghts.
To solve the radiative transfer problem we used the 1D-code of Manske \& Henning
(\cite{manske98})
assuming spherically symmetric shells like in previous works. 
As for the dust composition, we use graphite, silicate, and iron with optical
constants taken from Dorschner et al. (\cite{dorschner95}) and Draine \& Lee (\cite{draine84})
together with a
MRN-type size distribution (Mathis et al. \cite{mathis77}).
The model parameters and the related references are given in Tab.~\ref{par}: here 
L$_{*}$ is the stellar luminosity, R$_{in}$ and R$_{out}$  are
the inner and the outer radius of the envelope, and M$_{d}$ is the dust mass.
We underline that the inner radius of the dust shell of W3(OH) coincides with the outer
radius  of the \uchii{} while for the hot core, R$_{in}$ is derived from the dust
sublimation temperature ($\sim 1000$ K).
To model the W3(H$_2$O) SED we follow Osorio et al. (\cite{osorio99}) in
adding an accretion luminosity of $4.2\times10^4$ L$_{\odot}$ to the stellar luminosity 
and a free-fall density shell.
For W3(OH) the density profile that better fits our observations is gaussian  
and yields an almost constant density distribution till 
$\sim 20000$ AU. The resulting opacities at the mm wavelenghts varies as $\nu^{1.3}$.\\
% hot core opacity is 7.3 at 100\,\micron{}. The opacity  of W3(OH) at 0.55\,\micron{} is
%100.
Fig.~\ref{fig:sed} shows the individual SEDs as well as their superposition. 
Our model produces the following FWHMs at the observed wavelenghts: 1\farcs6 at 8.8\,\micron{},
2\farcs0 at 11.7\,\micron{} and 3\farcs1 at 17.9\,\micron{}. They are
quite in agreement with those given in Tab.~\ref{prop} which supports the view that
the parameters of the radiative transfer model are representative for the actual
conditions.
From the SED of W3(H$_2$O) we conclude that deep observations in the Q band might be
able to detect the MIR emission from the hot core.
%%%%%%%
%  here are the fluxes that i have from the model:
% W3(OH): F(8.8) = 4.180 Jy, F(11.7) = 45.10 Jy, F(17.9) =  59.30 Jy, F(21.4) = 235.0 Jy
% W3(H2O): F(1300) = 1.34 Jy, F(2600) = 0.139 Jy, F(3400) = 0.0221 Jy

  \begin{table}
      \caption[]{Radiative transfer model parameters}
              \label{par}
         \begin{tabular}{l l c l l l }
            \hline
           \noalign{\smallskip}
            Source  & L$_{*}$  & R$_{in}$& R$_{out}$&M$_{d}$   \\
	            &[$10^4$L$_{\odot}]$ & [AU]     & [AU]      & [M$_{\odot}$]  \\
           \noalign{\smallskip} 
            \hline
            \noalign{\smallskip}
            W3(OH)  &8$^{1}$ &2270$^{2,3}$  & 57000$^{4}$     & 5.0  \\
         W3(H$_2$O) &2.4$^{5,6}$ &190 & 24000$^{5}$ & 3.0$^{7}$\\
	       \noalign{\smallskip}
	   \hline 
         \multicolumn{6}{l}{ }\\
	 \multicolumn{6}{l}{Ref: (1) Wink et al. 1994, (2) Chini et al. 1986, }\\
	 \multicolumn{6}{l}{(3) Campbell et al. 1989, (4) Zeng et al. 1984, }\\
	 \multicolumn{6}{l}{(5) Osorio et al. 1999, (6) Thompson 1984,}\\
	 \multicolumn{6}{l}{(7) Wyrowski et al. 1997} \\
	   
	        \end{tabular}
   \end{table}  

%_______________________________________________________
\section{Summary}

   \begin{enumerate}
      \item The thermal infrared counterpart to the hot core source W3(H$_2$O) was not
       detected at any wavelength of our observations. This revises the finding
       of Keto et al. (\cite{keto92}) which most probably resulted from confusing W3(H$_2$O) with
       the \uchii{} northeast of W3(OH). We derived a lower limit for the extinction
       at 11.7\,\micron{} towards W3(H$_2$O) of 12\,mag.
       \item Our diffraction-limited imaging resolved the thermal emission from W3(OH)
       and clearly indicates a wavelength dependence of its apparent size.
       \item The comparison of our 8.8\,\micron{} flux to those measured with IRAS and MSX
       led to the conclusions that PAHs surround W3(OH).
       \item In accordance with our thermal infrared imaging and mm/submm studies, the
       ISO-LWS spectrum of W3(OH) might be decomposed in two components with W3(OH) being
       the hotter, more evolved, object while W3(H$_2$O) dominating at mid/far-infrared
       wavelengths. 
   \end{enumerate}

\begin{acknowledgements}
        We thank S. Rinehart for assistence during the observations and D. Wilner for
        providing the 8.4\,GHz VLA map.
       This work was supported by the
      \emph{Deut\-sche For\-schungs\-ge\-mein\-schaft} grant
      Ste~605/18--1.
\end{acknowledgements}

\end{document}